\documentclass[10pt, conference]{IEEEtran}

\usepackage[format=plain, font=sc, justification=centering,singlelinecheck=false]{caption}
\usepackage{array, color}
\newcolumntype{$}{>{\global\let\currentrowstyle\relax}}
\newcolumntype{^}{>{\currentrowstyle}}

\usepackage{verbatim}

\usepackage{enumitem}
\usepackage{soul}
\usepackage{color}
\usepackage{url}
\usepackage{xspace}
\usepackage{booktabs}
\usepackage{tabularx}

\newcommand\clearrow{\global\let\rowmac\relax}
\clearrow

\usepackage{multirow}
\usepackage{anyfontsize}
\usepackage{balance}
\usepackage{graphicx}
\usepackage{microtype}
\usepackage{pslatex}
\usepackage{seqsplit}
\usepackage{amssymb}
\usepackage{amsmath}
\usepackage{algorithmic}
\usepackage[ruled,vlined,procnumbered]{algorithm2e}
\usepackage[table,xcdraw]{xcolor}
\usepackage{textcomp}
\usepackage{multirow, hhline}
\usepackage{subfig}

\usepackage{tikz}

\usepackage{hyperref}

\newcommand{\ACDC}{\emph{ACDC}\xspace}
\newcommand{\ARC}{\emph{ARC}\xspace}

\newcommand{\PKG}{\emph{PKG}\xspace}

\definecolor{brown}{cmyk}{0,0.81,1,0.60}
\definecolor{magenta}{rgb}{0.4,0.7,0}
\definecolor{gray}{rgb}{0.5,0.5,0.5}
\definecolor{red}{rgb}{1,0,0}
\definecolor{yellow}{rgb}{.85,0.75,0}
\definecolor{purple}{rgb}{0.5,0,0.5}
\definecolor{green}{rgb}{0.0,0.2,0.0}
\definecolor{blue}{rgb}{0,0,1}

\definecolor{lgray}{gray}{0.85}
\definecolor{dgray}{gray}{0.65}

\newcolumntype{L}[1]{>{\raggedright\let\newline\\\arraybackslash\hspace{-4pt}}m{#1}}
\newcolumntype{C}[1]{>{\centering\let\newline\\\arraybackslash\hspace{-5pt}}m{#1}}
\newcolumntype{R}[1]{>{\raggedright\let\newline\\\arraybackslash\hspace{-5pt}}m{#1}}

\makeatletter
\def\@copyrightspace{\relax}
\makeatother

\begin{document}
%

\title{Architectural Decay as Predictor of\vspace{-0.5mm} \\ Issue- and Change-Proneness\vspace{-2mm}}

\author{
%

\IEEEauthorblockA{Duc Minh  Le*, Suhrid Karthik$^\dagger$, Marcelo Schmitt Laser$^\dagger$, and Nenad Medvidovic$^\dagger$}
\setlength{\tabcolsep}{40pt}
\begin{tabular}{ c c }
Software  Infrastructure*  &Computer Science Department$^\dagger$  \\
 Bloomberg L.P.  & University of Southern California \\
London, EC4N 4TQ, UK & Los Angeles, CA 90089, USA \\
dle50@bloomberg.net & \{skarthik,schmittl,neno\}@usc.edu \\ 
\end{tabular}
}

\date{19 Feb  2021}

\maketitle

\begin{abstract}

Architectural decay imposes real costs in terms of developer effort, system correctness, and performance. Over time, those problems are likely to be revealed as explicit implementation issues (defects, feature changes, etc.).  Recent empirical studies have demonstrated that there is a significant correlation between architectural ``smells''---manifestations of architectural decay---and implementation issues. In this paper, we take a step further in exploring this phenomenon. 
We analyze the available development data from 10 open-source software systems and show that information regarding \emph{current} architectural decay in these systems can be used to build models that accurately predict \emph{future} issue-proneness and change-proneness of the systems' implementations.  As a less intuitive result, we also show that, in cases where historical data for a system is unavailable, such data from other, unrelated systems can provide reasonably accurate issue- and change-proneness prediction capabilities. 

\end{abstract}



\section{Introduction}
\label{sec:introduction} 

\looseness-1
Software systems change regularly, as do their architectures. 
Over time, a system's architecture is increasingly affected by  decay, caused by careless or unintended 
 design decisions~\cite{perry1992foundations}. Decay results in systems whose implemented architectures differ in important ways 
 from their designed architectures. 
Both researchers and practitioners have recognized the negative impact of architectural decay and its role in causing technical debt. Despite this, when developers modify a system during maintenance, they often  focus on code and  neglect the  architecture. 


\looseness-1
Researchers have proposed a  number of  techniques to analyze a  system at the code level 
and to predict  issues that are likely to appear in the system's future versions. 
A common approach  has been to use historical artifacts, such as data from issue trackers and version control systems, to build prediction models. 
Early approaches \cite{Kim:2007:PFC:1248820.1248881, Moser:2008:CAE:1368088.1368114, Gyimothy2005, Coleman1994}  built models to predict implementation issues based on code metrics.
Later studies  made use of other 
 properties that were reckoned to be potential causes of issues, such as code dependencies 
~\cite{Zimmermann:2008:PDU:1368088.1368161} 
and code smells~\cite{Hall:2014:CSS:2668018.2629648}. 

\looseness-1
In contrast to code-level techniques, analogous techniques at the architecture level have not received nearly as much attention, even though recent  work has demonstrated that even  simple code updates can cause system-wide architectural changes~\cite{leempirical}. Frequently, such updates introduce \emph{architectural smells} in a system (e.g., dependency cycle, ambiguous interface~\cite{icsa2018duc}). These smells may have no immediately visible effect, but they are symptoms of architectural decay and accumulated technical debt~\cite{taylor2009software, joshthesis2014, leempirical, icsa2018duc}. 
As  decay  compounds in long-lived systems, the number of architectural smells grows, creating unforeseen issues when engineers try to modify a system. 

In such cases, engineers are eventually likely to realize the negative effects of the incurred technical debt and the need to refactor their system. However, they usually spot  deeper architectural problems only when related implementation-level issues surface. For example,  issue \#1178 reported for Apache Pig indicates that developers recognize the problem of having a large number of functions in a component: ``[The component] has been an area of numerous bugs, many of which have been difficult to fix''~\cite{pig-1178}. Similarly,  issue \#223 in Apache CXF acknowledges the need to refactor CXF's  architecture to reduce the amount of code changes and improve extensibility~\cite{cxf-223}.  

\begin{figure*}[tbh]
	\centering
	\includegraphics[width=18cm]{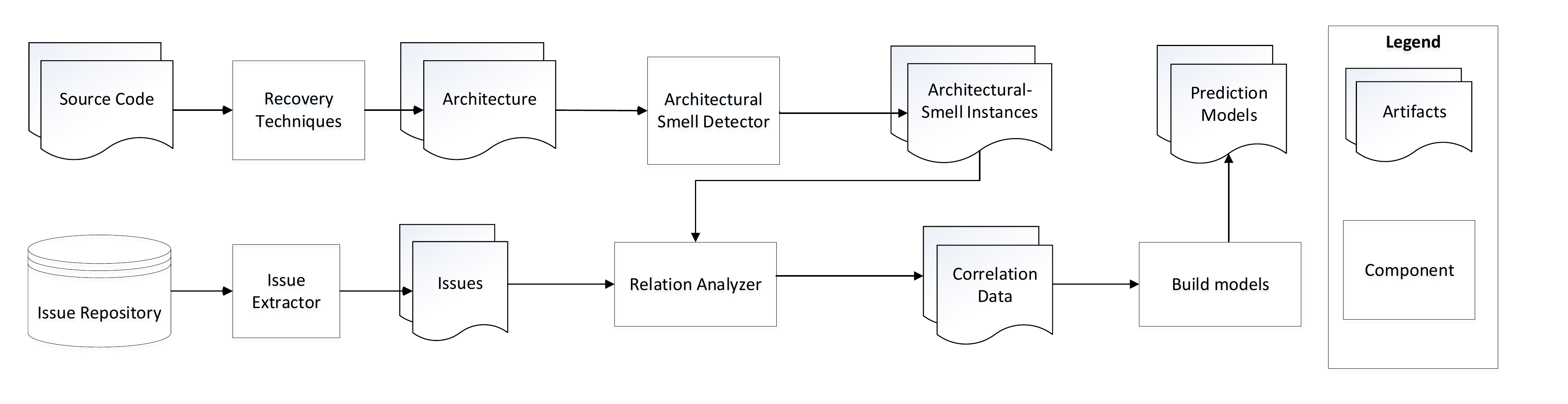}
	\vspace{-5mm}
	\caption{Architecture recovery pipeline used in our study and enabled by the ARCADE tool suite.}
	\label{fig:workflow}
	\vspace{-5mm}
\end{figure*}

Recent studies have established strong correlations between architectural smells and both (1) a system's proneness to change and (2)  the emergence of certain implementation issues~\cite{icsa2018duc, le2016relating}.  Furthermore, many bugs reported for a system have been shown to have architectural roots~\cite{Xiao:2014:DPA:2635868.2661679, 10.1007/978-3-030-00761-4_21}.   
Prior work has also demonstrated that identifying \emph{code} smells using existing approaches will not help to uncover  the underlying architectural issues, and modifications to address thus identified problems run the risk of being inadequate, short-term patches~\cite{OizumiSBES6943486}.
Despite this, predictive models 
that leverage \emph{architectural characteristics} to anticipate the implementation issues or the amount of change a system may experience have been scarce.


In this paper, we propose and empirically evaluate an approach to predict a system's (1)~future implementation issues and (2)~proneness to change based on the system's current and past architectural characteristics. 
Our work is inspired in part by the recent  finding~\cite{icsa2018duc} that architectural smells and implementation issues are strongly correlated.
Specifically, we analyze 466 versions of 10 open-source software systems. For each system version, we use 3 different methods 
to recover its architectures from source code. We analyze thus obtained 1,398 architectural models to detect 11 distinct types of architectural smells. The detected smells are subsequently used as features in our prediction models.
We make use of different machine learning techniques 
to predict a given system's issue- and change-proneness based on the collected architectural features. 

Our study has resulted in two principal findings regarding the predictive power of the models obtained in this manner: 

\begin{enumerate}
	\item 
	The architectural smells detected in a system can help to accurately predict both the issue-proneness and change-proneness of that system at a given point in time. 
	Our models yielded precision and recall scores of at least 70\% (and as high as 95\%) for specific recovered architectural views of the subject systems. 
	This finding allows maintainers to foresee future problems involving new smell-impacted parts of a system. 

	\item Different, independently developed software systems tend to share issue- and change-proneness characteristics.
	This allows developers to use 
	models created  using data from a set of existing  systems to predict the issue- and change-proneness of an unrelated system for which historical data does not exist (e.g., a newly developed system). 
	While the accuracy of such general-purpose prediction models is lower than the system-specific models, the loss in accuracy is moderate, typically under 10\%. Our results indicate that this is a fruitful area for further investigation, and that our models are already usable in practice for making certain types of decisions. 
\end{enumerate}

Section \ref{sec:foundation}  introduces foundations for our study. Section \ref{sec:exp_setup} presents the research questions and describes the study. The results 
are detailed in Section \ref{sec:results}. Threats to validity, 
related work,  conclusions, and acknowledgment round out the paper.


\bgroup
\def\arraystretch{1.15}
\begin{table*}[t]
	\centering
	\caption{Consolidated catalog of architectural smells}
	\label{tab:catalog_of_smells}
	\begin{tabular}{|p{1.35cm}|p{2.8cm}|p{6.92cm}|p{5.55cm}|}
		\hline
		\rowcolor[HTML]{C0C0C0} 
		\begin{tabular}[c]{@{}l@{}}Category\end{tabular} & \begin{tabular}[c]{@{}l@{}}Type\end{tabular} & Definition & Consequences   \\ \hline
		
		\multirow{5}{1.5cm}{Interface-based} & Unused Interface                         & Component's interface is not linked to other components                                                                             & Adds unnecessary complexity to the system                                                                      \\ \cline{2-4} 
		& Unused Brick                      & Component's interfaces are all unused                                                                             & Same as Unused Interface, but more severe                                                                    \\ \cline{2-4} 
		& Sloppy Delegation                                     & Component delegates functionality  it could have performed 
		& Reduces separation of concerns                                                                               \\ \cline{2-4} 
		&  \mbox{Functionality} Overload                          & Component has an excessive amount of functionality                                                                                                                                                                                                             & Reduced modularity                                                                        \\ \cline{2-4} 
		& Lego Syndrome                                         & Component handles  exceedingly small amount of functionality                                                                                                                                                                                                        & High coupling                                                                           \\ \hline
		\multirow{2}{1.5cm}{Change-based}& Duplicate \mbox{Functionality  }                             & Several components replicate the same functionality                                                                                                                                                                                               & Bugs if changing only one duplicate                                \\ \cline{2-4} 
		& Logical Coupling                                      & Parts of different components are frequently changed together                                                                                                                      & Similar  to Duplicate Functionality                                                                                                          \\ \hline
		\multirow{2}{1.5cm}{Dependency-based}& Dependency   ~~~  Cycle                                      & Set of components whose links form a circular chain                                                                                                                                                                                                                   & Changes to one component affect the entire cycle                             \\ \cline{2-4} 
		& Link Overload                                         & Component's interfaces have too many dependencies                                  & Reduced isolation of changes                                                                                                               \\ \hline
		\multirow{2}{1.5cm}{Concern-based}  & Scattered Parasitic Funct.                     & Multiple components  responsible for realizing one concern                                                                                                                                                                         & Changing a feature modifies multiple system parts \\ \cline{2-4} 
		& Concern Overload                                      & Component implements an excessive number of concerns                                                                                                                                                                                                             &  Violates separation of concerns                                                                                             \\ \hline
		
	\end{tabular}
	\vspace{-3mm}
\end{table*}
\egroup

\section{Foundation}
\label{sec:foundation}

Our work is directly enabled by three research threads: (1)~software architecture recovery, (2)  definition and analysis of architectural smells, and (3) tracking implementation issues. Figure~\ref{fig:workflow} depicts how these threads are combined to answer our research questions in this paper.

\subsection{Architecture Recovery with ARCADE}
\label{subsec:arcade}

Garcia et al.~\cite{garcia2013comparative}  conducted a comparative evaluation of software architecture recovery techniques. Their objective was to measure the existing techniques' accuracy and scalability on a set of systems for which researchers had previously obtained ``ground-truth'' architectures~\cite{garcia2013obtaining}. To that end, the authors implemented a tool suite, named ARCADE, offering a large set of architecture recovery choices to an engineer.\footnote{The existing   techniques implemented within ARCADE support  structural clusterings of software systems' elements based on a range of criteria. While the resulting recovered models contain only partial architectural information for a given system, in this paper we will refer to them as ``recovered architectures''. We note that our use of this term is consistent with existing literature.} 

Garcia et al.'s results indicate that two techniques implemented in ARCADE 
consistently outperformed the rest: \ACDC~\cite{tzerpos2000} and \ARC~\cite{garcia2011enhancing}. 
We select these techniques for our study. 
%
\ACDC leverages a system's \emph{structural characteristics} to cluster implementation-level modules into architectural components, while \ARC focuses on the \emph{concerns} implemented by a system. \ACDC relies on static dependency analysis; \ARC uses information retrieval and machine learning. 

\looseness-1
 \PKG is another technique implemented in ARCADE. \PKG extracts a system's implementation \emph{package structure}. The package structure of a system is considered to be a reliable view of a system's ``implementation architecture''~\cite{kruchten1995}. We use it to complement the two selected clustering-based architectural views.  

\subsection{Architectural Smells}
\label{subsec:cat_of_arch_smells}

Architectural smells are 
instances of poor architectural design decisions \cite{mens2004}. They negatively impact  system lifecycle properties, such as understandability, testability, extensibility, and reusability \cite{garcia2009toward}.
While code smells \cite{fowler1999}, anti-patterns \cite{buschmann2007pattern}, or structural design smells \cite{ganesh2013towards} originate from implementation constructs (e.g., classes, methods, variables), architectural smells stem from poor use of software architecture-level abstractions --- components, connectors, interfaces, patterns,  styles, etc. 
Detected instances of architectural smells  
are candidates for restructuring \cite{bowman1999}, to help prevent  architectural decay and improve system quality.

Researchers have collected a growing catalog of architectural smells. Garcia et al. \cite{garcia2009toward,garcia2009identifying}  identified an initial set of four  smells related to connectors, interfaces, and concerns. Mo et al.~\cite{mo2013mapping} introduced a new concern-related smell. Ganesh et al. \cite{ganesh2013towards} also summarized a catalog of structural design smells, some of which are at the architecture-level. Le et al.~\cite{icsa2018duc} described 11 different  architectural smells and proposed a set of algorithms to detect them. Table~\ref{tab:catalog_of_smells} summarizes a consolidated list of smells that were identified in the above references, after removing duplicates and non-architectural smells.

\subsection{Issue Tracking Systems}
\label{subsec:jira}

Issue tracking systems are commonly used  development tools that allow users to report different problems and concerns about a system and monitor their status. 
All subject systems selected for analysis in this paper use Jira~\cite{apachejira} as their issue tracking system. However, this is not a limitation; our approach  can be applied to other issue trackers. 

\looseness-1
When reporting implementation issues, engineers categorize them into different  types: \emph{bug}, \emph{new feature}, \emph{improvement}, \emph{task}  to be performed, etc. We consider all issue types in our study because they may result in relevant changes to a system. In other words, any issue type or individual issue instance may have an underlying architectural cause. Note that it would be possible to perform a finer-grained analysis using the same process we employed that would focus on a specific subset of issues or types. 

Each issue has a status that indicates where the issue is in its lifecycle~\cite{jiraissue}. An issue starts as ``open'', progresses to ``resolved'', 
and finally to  ``closed''. 
We restrict our study to closed and resolved issues that have been ``fixed'', and ignore those resolved issues that fall under ``won't fix'', ``cannot reproduce'', etc. We do so because 
any effects caused by the fixed issues presumably appear in certain system versions and disappear once the issue is addressed. Additionally, a fixed issue contains information that is useful for our study:
(1)~\emph{affected versions} in which the issue has been found,
(2)~\emph{type} of  issue, and
(3)~\emph{fixing commits}, i.e., the changes applied to the system to resolve the issue. 
Finding fixing commits is not always easy since there is no standard method for  engineers to keep track of this information. 
Three ways of keeping track of an issue's fixing commits  are commonly employed in our set of subject systems: (1) direct links to the commits, (2) specifying pull requests, and (3)~specifying patch files. Our implemented tool supports collecting data from all three methods.

Based on the collected information, issues are mapped to detected smells. 
To do this, first, we find the system versions that the issue affects. Then we find the architectural smells present in those versions. We say the issue is infected by a given smell if and only if (1) both the issue and the smell affect the same system version and (2) the resolution of the issue changes files that are involved in the smell. Based on this relationship, we studied if the characteristics of an issue (e.g. issue type, number of fixing commits) depend on whether the issue is infected by a given smell. 

\looseness-1
Note that resolving an issue may not remove the smell that led to the issue in the first place. One reason is that  developers could find a workaround. The smell may also~correlate with more than one issue. In general, it is difficult to identify the exact relationship between a specific architectural smell instance and a specific implementation issue. Fortunately, we do not need to do that in our work, because we are looking for prediction models that  uncover smell-issue correlations across most cases.

\section{Empirical Study Setup}
\label{sec:exp_setup}

This section describes our study setup. Our hypothesis and research questions are described in Section \ref{subsec:Research_questions}. 
 We then describe how we pre-processed the raw data in Section~\ref{subsec:extending_arcade}.

\subsection{Research Question and Research Hypothesis}
\label{subsec:Research_questions}


\begin{figure*}[t]
	\centering
	\includegraphics[scale=0.5]{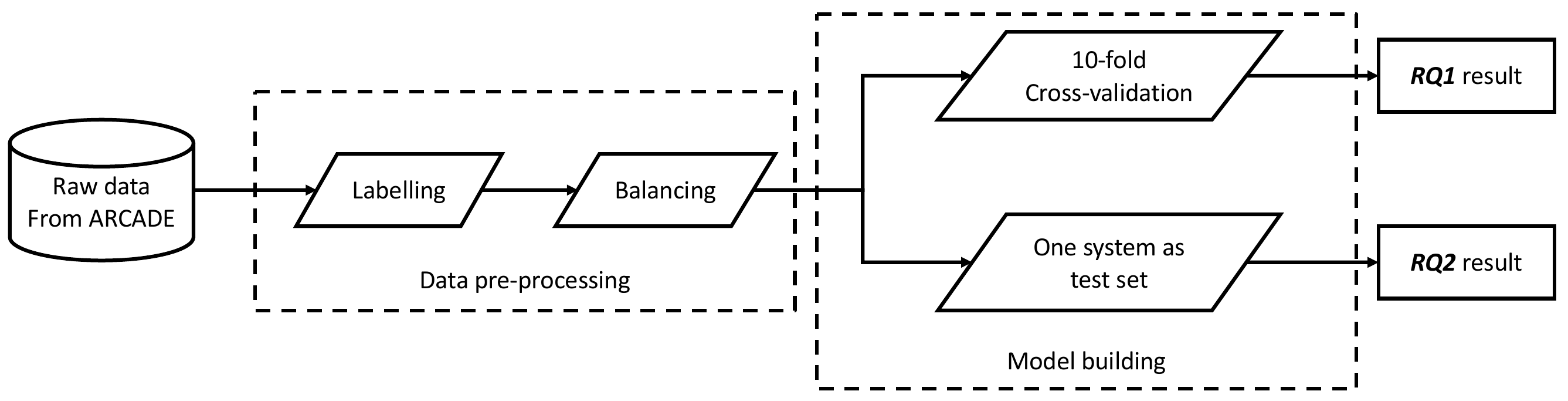}
		\vspace{-2mm}
	\caption{Data processing pipeline.}
		\vspace{-5mm}
	\label{fig:datapipe}
\end{figure*}

Our \textit{\textbf{hypothesis}} is that \emph{it is possible to construct accurate models to predict the impact of architectural decay on a system's implementation}.
To evaluate this hypothesis, we focus on 
the predictability of a system's issue- and change-proneness based on the identified architectural smells (i.e., the symptoms of decay). 
We define two research questions accordingly.

\vspace{1.5mm}
\textit{\textbf{RQ1.}} \emph{To what extent can the architectural smells detected in a system help to predict the issue-proneness and change-proneness of that system at a given point in time?}
\vspace{1.5mm}

The training data used to build the prediction models for a system is collected from different versions of that system. 
If these models can be shown to accurately predict issue- and change-proneness, this would indicate that architectural smells have consistent impacts on those two properties throughout a system's life span. In turn, this would confirm that the impact of architectural smells is not related to other factors, such as system size, which will change during a system's evolution. In addition, an accurate prediction model will be useful for maintainers to foresee the future issue- and change-proneness of newly smell-affected parts of a system, helping them 
to decide when and where they may need to refactor the system. 

\vspace{1.5mm}
\textit{\textbf{RQ2.}} \emph{To what extent do unrelated software systems tend to share properties with respect to issue-proneness and change-proneness?}
\vspace{1.5mm}

\looseness-1
This question investigates whether the issue- and change-proneness of a system can be accurately predicted by a general-purpose model trained using symptoms of architectural decay from unrelated  systems. If such a model can be constructed,  it can be reused by developers to predict properties of systems for which historical information is not (yet) available. 
An affirmative answer to this question would also have a deeper implication: software systems tend to share fundamental properties regardless of system type, application domain, developers, employed tools, programming languages, execution platforms, etc.

\subsection{Building the Data Pipeline}
\label{subsec:extending_arcade}

\looseness-1
To answer the  two research questions, we
build multiple prediction models based on different systems' architectural-smell data and  assess the models' accuracy. We rely on ARCADE~\cite{leempirical} to collect the underlying raw architectural-smell data, and WEKA\cite{weka}---a well-known ML framework---to pre-process the data, build prediction models, and evaluate their accuracy. The data pipeline we use is illustrated in Figure \ref{fig:datapipe}. Section \ref{subsec:subjects} introduces the list of subject systems and the process of recovering their architectural artifacts with ARCADE. Two main pre-processing tasks 
are labeling and balancing the raw data, which are discussed in Sections \ref{sec:labeling} and \ref{sec:balancing}, respectively. Creating the training and test sets,  evaluating prediction models  as well as determining the baseline models are discussed in Sections \ref{sec:training_test},  \ref{sec:metrics}, and \ref{sec:baseline}, respectively. 

\bgroup
\def\arraystretch{1.15}
\begin{table}[b]
	\vspace{3.25mm}
	\centering
	\caption{Subject systems in our study}
	\begin{tabular}{|l|p{2.45cm}|p{1.2cm}|p{0.9cm}|p{1.2cm}|}
		\hline
		\rowcolor[HTML]{C0C0C0} 
		{System} & {Domain}   & {{\# Versions}} & {\# Issues} & {Avg. LOC} \\	\hline
		Camel       & Integration F-work  & 78             & 9665      & 1.13M \\ 		\hline      
		CXF         & Service F-work      & 120            & 6371      & 915K  \\		\hline
		Hadoop      & Data Proc. F-work & 63             & 9381      & 1.96M \\		\hline
		Ignite   	&  In-memory F-work 	 &  17   		  &  3410		  & 1.40M \\		\hline
		Nutch       & Web Crawler      		 & 21             &	1928	      & 118K  \\		\hline
		OpenJPA		& Java Persist. 	 	 & 20    		  & 1937 	  & 511K \\		\hline
		Pig   		&  Data Analysis F-work     &            16  &  3465     & 358K  \\ 		\hline
		Struts2     & Web App F-work      & 36             & 4207      & 379K  \\		\hline
		Wicket      & Web App F-work      & 72             & 6098      & 332K  \\		\hline
		ZooKeeper   & Config. Mgmt F-work    &  23            &   1390    & 144K  \\		\hline
	\end{tabular}%
	\label{tab:subject_systems_predict}%
\end{table}%
\egroup

\vspace{1mm}\subsubsection{ARCADE and Subject Systems}
\label{subsec:subjects}
We collected  data  from ten open-source systems from the Apache Software Foundation, shown in Table~\ref{tab:subject_systems_predict}. 
Specifically, our study uses three types of data: (1) architectural smells detected in recovered architectures, (2) implementation issues collected from the Jira~\cite{apachejira} issue repository, and (3) code commits extracted from GitHub \cite{github}. 

Using ARCADE, we recover the subject systems' architectures using the three recovery techniques---ACDC, ARC, and PKG---whose accuracy and scalability have been demonstrated by prior work (recall Section~\ref{subsec:arcade}). We then analyze the recovered architectures for the presence of smells  identified in the literature (recall Section~\ref{subsec:cat_of_arch_smells} and Table~\ref{tab:catalog_of_smells}), as well as the systems' issue- and change-proneness. 
Those architectural artifacts are the raw data for building prediction models.

\vspace{1mm}\subsubsection{Labeling the Data}
\label{sec:labeling}
Data labeling is a key step to ensure the success of prediction models. 
In our prediction problem, we are interested in two properties---issue-proneness and change-proneness. These properties can be obtained by, first, counting the raw numbers of issues and changes in a system's development history and, then, finding a way of characterizing those numbers. 
Specifically, we assign nominal labels based on the raw numbers of issues and changes related to source files to represent different levels of issue- and change-proneness.

\begin{figure}[b]
		\vspace{2mm}
	\centering
	\includegraphics[scale=0.52]{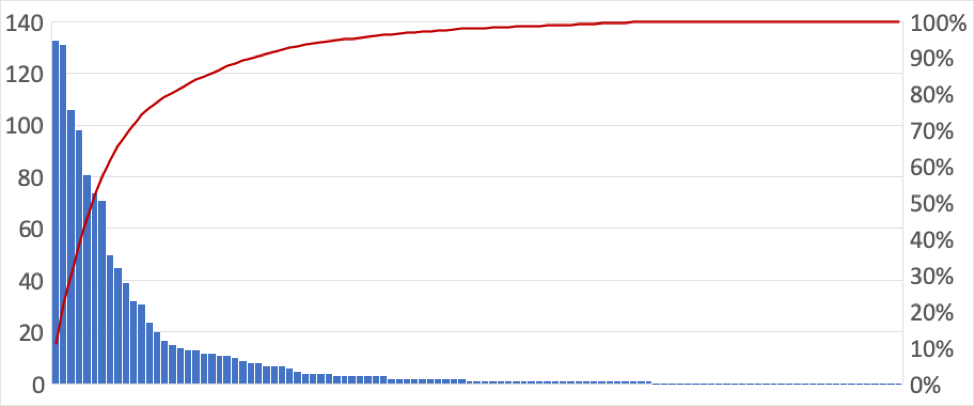}
	\caption{Pareto chart of issues per file in Hadoop. The x-axis represents the Hadoop files grouped by the number of issues they contain, the left y-axis  the number of files in same groups, and the right y-axis  the cumulative percentage of groups' sizes.} 
	\label{fig:long-tail}
\end{figure}

\looseness-1
Converting a set of numerical values to nominal labels~depends on the  values' distribution. In our problem, the~numbers of  issues and changes follow a heavy-tailed distribution~\cite{foss2011introduction}, where many files  are associated with small numbers of issues and code changes, while comparatively fewer files are associated with large numbers of issues and  changes. This is not an uncommon type of distribution~\cite{yamashita2015revisiting, boehm2006value}.  
As an illustration, the Pareto chart \cite{doi:10.1198/000313006X152243} in Figure \ref{fig:long-tail} depicts the distribution of issues per file in Hadoop: 
while few files  are associated with a large number of issues, the arc, which represents the cumulative percentage of file-groups' sizes, shows a clear heavy-tailed pattern.

\begin{figure*}[t]
	\centering
	\includegraphics[scale=0.58]{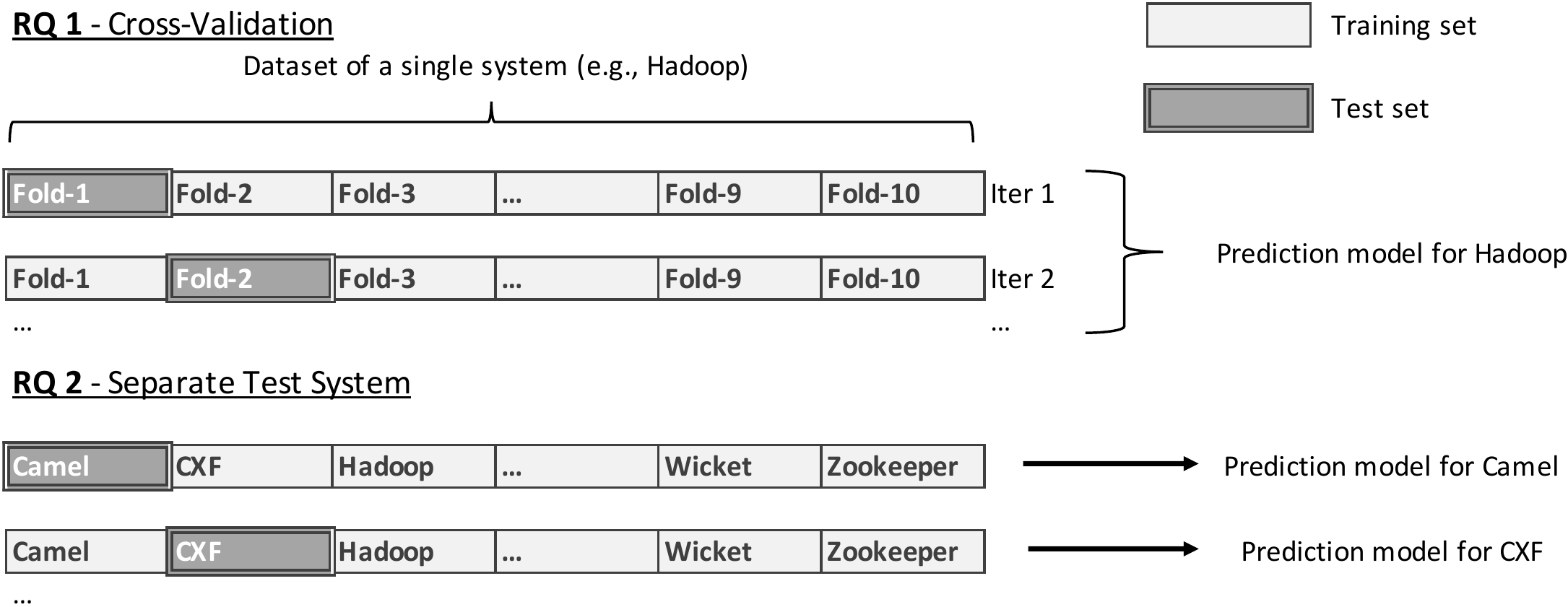}
	\vspace{-2mm}
	\caption{Creating datasets to answer RQ1 (top) and RQ2 (bottom).} 
	\vspace{-5mm}
	\label{fig:divide_sets}
\end{figure*}

One common labeling approach is to segment a heavy-tailed distribution into head and tail segments. A more sophisticated approach is to divide the distribution into 
three parts---head, body, and tail---which in our case represent the three levels of proneness: low, medium, and high. We  choose the latter approach because the numerical values in our study span a wide range. Having these three levels gives developers a better estimation of architectural decay's impact.
 
 To segment a dataset, we use the Pareto principle \cite{pareto1906manuale}, a popular segmentation method for heavy-tailed distributions,  widely used in software engineering (e.g., \cite{boehm2006value, kiremire2011application, sayyad2013pareto}). To obtain the three segments, we apply the Pareto principle twice, 
as suggested in  literature \cite{arthur2001six}. Specifically,  we divide the original dataset into two portions. The first portion contains 80\% of the original dataset's low-end, while the second portion contains 20\% of the high-end. We apply the Pareto segmentation once more to the latter portion, thus obtaining two new portions that respectively contain the next 16\% (80\% of the 20\%) and 4\% (20\% of the 20\%) of the high-end data points. 

In order to collect the data regarding architectural decay, 
for each version of a subject system, we first collect the list of  ``fixed'' issues affecting that version. 
Next, we collect the files that were changed when fixing the issues. For each file, we gather its associated architectural smells, 
 the number of issues whose fixing commits changed that file (used when determining the system's \emph{issue-proneness} in Sections~{IV-RQ1-A}~and IV-RQ2-A), and the total number of changes (used when determining the system's \emph{change-proneness} in Sections~IV-RQ1-B and IV-RQ2-B). 
After the raw data is collected, we label it using the Pareto technique mentioned above 
before feeding it to supervised ML algorithms. 

To determine the level of issue-proneness of a source file in a system version, first, the number of issues related to that  file 
is collected. This is one data point. We collect data points for all files in all available versions of a  system, and then sort the dataset by the numbers of issues, from low to high.
Then, the first 80\% of data points are marked with ``low'' labels; the next 16\% and 4\%, respectively, are marked with ``med(ium)'' and  ``high'' labels. To determine the change-proneness of a source file in a system's version, we count the number of commits related to that file and repeat a similar labeling process.

\begin{table}[b!]
\vspace{2mm}
	\centering
	\caption{Data samples from Hadoop}
		\vspace{-1mm}
	\label{tab:data_point_sample}
	\setlength\tabcolsep{4.75pt}
	\begin{tabular}{|l|l|l|l|l|l|l|p{0.2cm}|p{0.352cm}|}
		\hline
		\rowcolor[HTML]{C0C0C0}
				Vers. &  Filename        & CO & SF & LO & DC & ... & Iss & Chg \\ \hline
		0.20.0 & dfs/DFSClient.java & 0                & 1   & 1             & 1                & ... & H            & L              \\ \hline
		0.20.0 & mapred/JobTracker.java         & 1                & 0   & 1             & 0                & ... & M             & M              \\ \hline
		0.20.0 & tools/Logalyzer.java       & 0                & 0   & 0             & 0                & ... & L             & L              \\
		\hline      
		... & ...       & ...                & ...   & ...             & ...                & ... & ...             & ...              \\
		\hline      
	\end{tabular}
\end{table}

Table \ref{tab:data_point_sample} shows several data samples in our datasets after labeling. 
The  shown features, i.e., architectural smells in our case, 
are CO (Concern Overload), SF (Scattered parasitic Functionality), LO (Link Overload), and DC (Dependency Cycle). The output features, i.e., labels, are the levels of issue-proneness  and change-proneness.  
The two leftmost columns show the versions and filenames of each data point. The next eleven columns are binary features that indicate the presence (1) or absence (0) of a specific smell (recall Table~\ref{tab:catalog_of_smells}) in a given file. 
The two rightmost columns indicate the issue-proneness  (``Iss'') and change-proneness (``Chg'') of the files. For example, in version 0.20.0 of Hadoop, \texttt{DFSClient.java} has three smells: SPF, LO, and DC. The file's issue-proneness is high (H), and its change-proneness is low (L). On the other hand, both issue- and change-proneness of \texttt{JobTracker.java} are medium (M). 

\vspace{1mm}\subsubsection{Balancing the Data}
\label{sec:balancing}
\looseness-1
Due to the distribution of data and the labeling approach, we need to balance our datasets~\cite{provost2000machine}. 
Recall from Section \ref{sec:labeling} that 
the {{$low:med:high$}} ratio of our datasets is 80:16:4 (i.e., 20:4:1).  If such a dataset were used to train a prediction model, 
the most likely outcome would be a model that predicts ``low'' for every data point. 
As we are more  interested in ``high'' and ``med'' labels, such a  model would be useless.
It is thus important to ensure that weighted metrics are not biased by less (or more) frequent labels.

We use SMOTE \cite{chawla2002smote} to balance our dataset, oversampling
``med'' by a factor of 5 and ``high'' by a factor of 20. 
SMOTE is a technique that synthesizes new minority samples based on nearest neighbors between sample data points. Adding new minority samples guarantees that the dataset will be balanced, i.e., that the {{$low:med:high$}} ratio will be 1:1:1.

\vspace{1mm}\subsubsection{Training and Test Sets}
\label{sec:training_test}
To build and test our prediction models, we use two different approaches for the two research questions, as illustrated in Figure \ref{fig:divide_sets}. In the first approach, used for RQ1, one dataset is created for each subject system with a cross-validation setup. Specifically, we use 10-fold cross-validation, where the dataset is randomly divided into ten equal-sized subsets. Then, we sequentially select one subset and test it against the prediction model built by the other nine subsets. The final result is the mean of the ten tests' results.  
In the second  approach, used for RQ2, we combine all subject systems and then divide them into two independent datasets: a training set, which comprises nine systems, and a test set, which comprises the single remaining system.

\vspace{2mm}\subsubsection{Evaluation Metrics}
\label{sec:metrics}
\looseness-1 
To evaluate the accuracy of our models, we use precision and recall
~\cite{powers2011evaluation}. Precision is the fraction of correctly predicted labels over all predicted labels. Recall is the fraction of correctly predicted labels over all actual labels. 

\looseness-1
For illustration, consider the sample confusion matrix, shown in Table \ref{tab:multiclass}, that is produced after classifying 25 samples into ``high'', ``med'', and ``low''. The precision for the ``high''  label is the number of correctly predicted ``high'' samples (4) out of all samples predicted to be ``high'' (4+3+6=13), i.e., $30.8$\%; its recall is the number of correctly predicted ``high'' samples (4) out of the number of actual ``high'' samples (4+1+1=6), i.e., $66.7$\%. We can similarly calculate the precision and recall for ``med''  and ``low''. Finally, we compute the average values of all  labels.

If a model predicts the correct labels, we consider this  a true positive. On the other hand, if the model predicts any of the three labels (``high'', ``med'', or ``low'') incorrectly, we consider this a false positive. This is the standard way of measuring the accuracy of multi-label problems \cite{tsoumakas2007multi}.

\begin{table}[t!]
	\centering	
	\caption{Example predicted vs. actual values}
	\vspace{-1mm}
	\begin{tabular}{ll|l|l|l|}
		
		\hhline{|~~|---|} 
		&      & \multicolumn{3}{c|}{\cellcolor[HTML]{C0C0C0}True/Actual}                                                 \\ \hhline{|~~|---|} 
		&      &High                       & Med                                             & Low                       \\ \hline
		\multicolumn{1}{|l|}{\cellcolor[HTML]{C0C0C0}}   & High  & \cellcolor[HTML]{EFEFEF}4 & 6    & 3                    \\ \hhline{|~|----|} 
		\multicolumn{1}{|l|}{\cellcolor[HTML]{C0C0C0}}   & Med & 1  & \cellcolor[HTML]{EFEFEF}2 & 0    \\  \hhline{|~|----|} 
		\multicolumn{1}{|l|}{\multirow{-3}{*}{\cellcolor[HTML]{C0C0C0}Predict}} & Low  & 1   & 2   & \cellcolor[HTML]{EFEFEF}6 \\ \hline
	\end{tabular}
\label{tab:multiclass}
\end{table}

\vspace{1mm}\subsubsection{Determining Baseline Models}
\label{sec:baseline}
To determine the effectiveness of the prediction models, we need to compare them to a baseline. In this case, we consider a baseline model to be  the simplest possible prediction. The model can be obtained through different approaches. For some problems this may be a random result, and for others in may be the most common prediction. 
As our dataset has been balanced (Section \ref{sec:balancing}), the simplest approach is ``uniform'' ---  generate predictions uniformly at random. This implies a prediction in which Table~\ref{tab:multiclass} has equal values in all cells, giving us a model with both precision and recall of 33.3\%.

\section{Empirical Study Results}
\label{sec:results}

In this section, for each of the two research questions we discuss the validation method and the associated findings. 

\vspace{3mm}
\noindent\emph{\textbf{RQ1:} To what extent can the architectural smells detected in a system help to predict the issue-proneness and change-proneness of that system at a given point in time?}
\vspace{1mm}

In this prediction problem, all  input features are binary (recall Table \ref{tab:data_point_sample}), indicating whether a file contains an architectural smell.
For this reason, decision-based techniques are most~likely to yield good results \cite{lim2000comparison}.  Metrics collected from a range of models we built and evaluated using four different classification techniques---decision table \cite{kohavi1995power}, decision tree \cite{quinlan2014c4}, logistic regression \cite{le1992ridge}, naive bayes \cite{john1995estimating}---confirmed this. 
We thus only discuss the  results obtained by 
the decision-table models. 

\vspace{3mm}
\noindent\emph{A. Issue-Proneness}
\vspace{1mm}

\looseness-1
Recall from Section~\ref{subsec:extending_arcade} that, to compute issue-proneness, for each file in each version of a given system, we gather the file's associated architectural smells and 
  number of issues whose fixing commits changed the file.  
Table \ref{RQ1.issues_proneness} shows the precision and recall of the models for predicting the issue-proneness of our subject systems from Table \ref{tab:subject_systems_predict}. These metrics are computed using 10-fold cross-validation~\cite{kohavi1995study}. 
The bottom-most row shows the average values across all systems. 
For each system, we built different prediction models based on smells detected in the three architectural views (ACDC, ARC, and PKG). In total, 30 prediction models per system were created and evaluated. 

\begin{table}[t]
	\centering
	\caption{Predicting issue-proneness}
	\label{RQ1.issues_proneness}
	\vspace{-1mm}
	\scriptsize
	\begin{tabular}{|l|c|c|c|c|c|c|}
		\hline 
		\rowcolor[HTML]{C0C0C0}		\multirow{2}{*}{} & \multicolumn{2}{c|}{ACDC}    & \multicolumn{2}{c|}{ARC}     & \multicolumn{2}{c|}{PKG}     \\ \cline{2-7} 
		\rowcolor[HTML]{C0C0C0}{System}		& Precision & Recall  & Precision & Recall  & Precision & Recall  \\ \hline
		Camel 					& 69.9\%    & 68.4\%  & 70.8\%    & 67.0\% &  68.2\%    & 62.8\%  \\ \hline
		CXF                     & 78.0\%    & 76.7\% &  68.9\%    & 68.3\%  & 64.7\%    & 63.8\%  \\ \hline
		Hadoop                  & 81.2\%    & 80.1\%   & 76.6\%    & 76.6\%   & 72.8\%    & 73.4\%     \\ \hline
		Ignite 					& 78.9\%	& 78.1\%	 & 78.9\%& 79.1\%& 70.4\%& 71.0\% \\ \hline
		Nutch                   & 80.8\%    & 71.6\%   & 82.5\%    & 82.7\%   & 68.3\%    & 52.1\%   \\ \hline
		OpenJPA                 & 71.4\%    & 68.3\%   & 74.5\%    & 73.2\%   & 69.2\%    & 67.9\%   \\ \hline
		Pig   					& 71.7\%	 &69.1\%	 & 71.3\%& 71.1\%& 68.6\%& 69.5\%\\ \hline
		Struts2                  & 89.2\%    & 89.0\%   & 95.0\%    & 94.8\%  & 79.1\%    & 78.3\%   \\ \hline
		Wicket                  & 69.2\%    & 70.1\%   & 76.7\%    & 77.1\%   & 63.7\%    & 65.4\%  \\ \hline
		ZooKeeper 	  	 & 72.0\%	&72.6\%	 & 70.8\%& 69.2\%&  68.7\%& 69.4\% \\ \hline
		Average   			& 76.2\%	&74.4\%	&76.6\%	&75.9\%	&69.4\%	&67.4\%	     \\ \hline
	\end{tabular}
	\vspace{1mm}
\end{table}

\begin{table}[b]
	\vspace{3mm}
	\centering
	\caption{Predicting issue-proneness with "high",  "med" and "low"  labels under ACDC }
	\label{RQ1.issues_proneness_high_med}
	\vspace{-1mm}
	\scriptsize
	\begin{tabular}{|l|c|c|c|c|c|c|}
		\hline 
		\rowcolor[HTML]{C0C0C0}		\multirow{2}{*}{} & \multicolumn{2}{c|}{High}    & \multicolumn{2}{c|}{Med}     & \multicolumn{2}{c|}{Low}     \\ \cline{2-7} 
		\rowcolor[HTML]{C0C0C0}{System}		& Precision & Recall  & Precision & Recall  & Precision & Recall  \\ \hline
		Camel 					& 73.9\%    & 56.9\%  & 57.6\%    & 63.6\% &  78.2\%    & 69.9\%  \\ \hline
		CXF                     & 94.4\%    & 83.3\% &  65.2\%    & 76.0\%  & 74.5\%    & 70.7\%  \\ \hline
		Hadoop                  & 71.2\%    & 81.5\%   & 78.3\%    & 78.1\%   & 72.1\%    & 81.5\%     \\ \hline
		Ignite 					& 93.8\%	& 89.1\%	 & 66.4\%& 76.8\%& 76.6\%& 67.8\% \\ \hline
		Nutch                   & 66.9\%    & 94.7\%   & 90.4\%    & 61.2\%   & 95.0\%    & 75.8\%   \\ \hline
		OpenJPA                 & 69.3\%    & 89.5\%   & 65.9\%    & 49.8\%   & 79.1\%    & 65.6\%   \\ \hline
		Pig   	 				   & 80.5\%	 &90.9\%	 & 72.9\%&  52.3\%&  61.8\%& 64.1\%\\ \hline
		Struts2                  & 96.3\%    & 95.7\%   & 88.2\%    & 81.1\%  & 83.1\%    & 90.4\%   \\ \hline
		Wicket                  & 78.8\%    & 89.6\%   & 59.3\%    & 60.3\%   & 69.5\%    & 57.3\%  \\ \hline
		ZooKeeper  		 & 71.0\%	&88.0\%	 & 64.2\%& 54.2\%&  80.7\%& 75.6\% \\ \hline
		Average   			& 79.6\%	&85.9\%	& 70.5\%	& 65.3\%	& 76.1\%	& 71.9\%	     \\ \hline
	\end{tabular}
\end{table}

\looseness-1
In general, the prediction models that relied on architectures recovered by ACDC and ARC were comparable in terms of accuracy: the average (precision, recall) for the ACDC and ARC models were  (76.2\%, 74.4\%) and (76.6\%, 75.9\%) respectively. On the other hand, the models emerging from PKG yielded accuracy that was up to 13\% lower. 
The models yielded very high predictive power in the cases of certain systems. For example, the ARC-based models for Struts2 achieved $\approx$95.0\% and the ACDC-based models  $\approx$90.0\% for each of the two metrics.

\begin{figure}[t!]
	\vspace{-4.5mm}
	\centering
	\subfloat[Precision]{{\includegraphics[width=4.3cm]{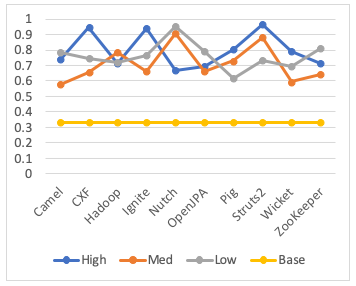} }}%
	\hspace{-0.9cm}
	\qquad
	\subfloat[Recall]{{\includegraphics[width=4.3cm]{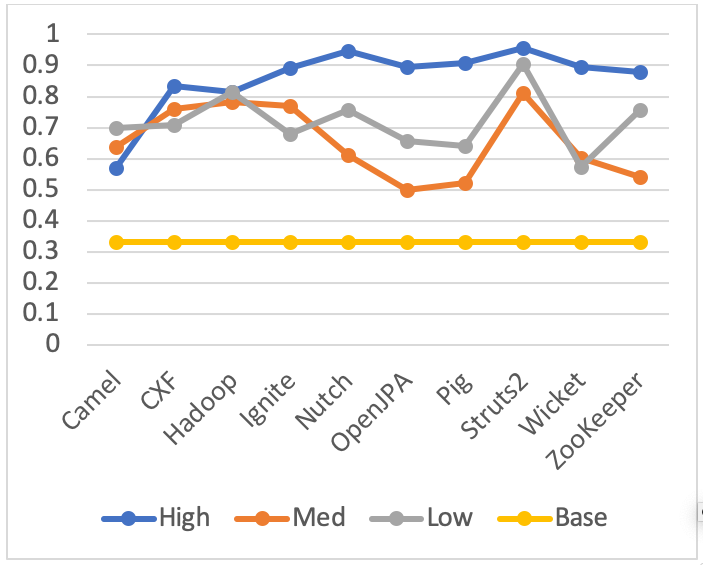} }}%
		\vspace{-1.5mm}
	\caption{Precision and Recall of issue-proneness prediction for each label in ACDC.}
	\label{fig:compare_rq1}
\end{figure}

	As discussed in Section \ref{sec:balancing}, our dataset has been balanced to ensure that  the trained models will accurately predict ``high'' and ``med'' labels, in which we are interested. Table \ref{RQ1.issues_proneness_high_med} shows the precision and recall of issue prediction for all three labels. While there are variations across the three labels, the average precision and recall for the ``high'' label---79.6\% and 85.9\%, respectively---outstrip the average values for the other two labels. Figure \ref{fig:compare_rq1} shows the comparison of our prediction models with the baseline model. Our prediction models are at least 1.5$\times$ better (2$\times$ in a majority of cases) than the baseline's 33.3\%, further confirming that our models are useful for predicting files with high numbers of related issues.

Our results confirm that architectural smell-based models can accurately predict the issue-proneness of a system. In other words, architectural smells have a consistent impact on a system's implementation with respect to issue-proneness over the system's lifetime. This finding means that \emph{architectural decay can be a powerful indicator of the health of a system's implementation}. It serves as a direct motivator for software engineers  to pay more attention to the architecture, and architectural smells,  in their systems. For example, system maintainers can use our models to foresee future problems, to devise refactoring plans, to prioritize their activities, etc. 

\looseness-1
The comparatively poorer performance of PKG in answering RQ1 suggests that implementation-package structure is not effective for measuring architectural decay and it can mask deeper architectural problems. This observation is in line with previous findings~\cite{leempirical}, which showed that, compared to ACDC and ARC, PKG  is markedly less useful for understanding the underlying architectural changes and their impact. 
 
This leads to another  observation. Recall the categorization of architectural smells in Section \ref{subsec:cat_of_arch_smells} and Table \ref{tab:catalog_of_smells}: two of the four categories are dependency-based and concern-based smells. This suggests that ACDC (dependency-based recovery) and ARC (concern-based recovery) should inherently outperform PKG when such smells are encountered. It further suggests that targeting specific recovery techniques to specific types of smells, and then finding a way to combine their results, may yield even higher accuracy in our prediction models. 
We are exploring this hypothesis in our ongoing work.

\vspace{3mm}
\noindent\emph{B. Change-Proneness}
\vspace{1mm}

Recall from Section~\ref{subsec:extending_arcade} that, to compute change-proneness, for each file in each version of a given system we gather (1)~the file's associated architectural smells and 
(2)~the total number of changes to the file reflected in the implementation issues' fixing commits. 
We used the same approach to evaluate the accuracy of the 30 architectural models for each system in predicting change-proneness as we did for predicting issue-proneness. 

\looseness-1
Table \ref{RQ1.change_proneness} shows the accuracy of our models. 
The models based on PKG-recovered architectures again have the lowest accuracy. In some systems, e.g., CXF and Nutch, the values for PKG-recovered architectures  are 10-20\% lower than the corresponding values in the other two views. 
The average (precision, recall) are (74.7\%, 71.6\%) and (73.6\%, 73.5\%)  for the ACDC- and ARC-based architectural views, respectively. Notably, the values yielded when analyzing Struts2 are, once again, very high.  
A further investigation of Struts2's dataset highlighted a distinguishing characteristic: 36 of the analyzed versions are distributed across just four minor Struts2 versions: 2.0.x, 2.1.x, 2.2.x and 2.3.x. In other words, the changes in most of these 36 versions were ``patches''. It is reasonable to expect that the architectures and detected smell instances between patches within a single minor version will be very similar. The prediction model for Struts2 benefits from this similarity and thus achieves very high accuracy in the cross-validation test. This suggests a promising strategy for building prediction models:  to increase the accuracy of models used to predict properties of a system version, one should \emph{select recent versions} instead of all versions across the entire system lifespan.


In summary, our results confirm that the historical data of a software system regarding its architectural smells, issues, and changes 
can be used to develop models to accurately predict the issue- and change-proneness of that system. The results also indicate that architectural smells have a consistent impact on software system implementations throughout the systems' lifetimes. 
Our architecture-based prediction approach, whose performance is usually two times better than the baseline, is useful for software maintainers to foresee likely future problems in newly smell-impacted parts of their system. The approach can also help in creating maintenance plans that can help to effectively reduce the system's issue- and change-proneness. Lastly, ACDC and ARC outperform PKG, emphasizing the importance of  selecting the appropriate architecture recovery techniques and targeting them to the task at hand. 


\begin{table}[t]
	\centering
	\caption{Predicting change-proneness}
	\vspace{-1mm}
	\scriptsize
	\label{RQ1.change_proneness}
	\begin{tabular}{|l|c|c|c|c|c|c|}
		\hline 
		\rowcolor[HTML]{C0C0C0}		\multirow{2}{*}{} & \multicolumn{2}{c|}{ACDC}    & \multicolumn{2}{c|}{ARC}     & \multicolumn{2}{c|}{PKG}     \\ \cline{2-7} 
		\rowcolor[HTML]{C0C0C0}{System}		& Precision & Recall  & Precision & Recall  & Precision & Recall  \\ \hline
		Camel                   & 69.9\%                         & 63.4\%                & 68.0\%                         & 67.1\%                                & 60.3\%                         & 61.0\%                                  \\ \hline
		CXF                     & 73.7\%                         & 70.8\%                 & 69.7\%                         & 63.4\%                       & 60.8\%                         & 63.4\%                     \\ \hline
		Hadoop                  & 78.1\%                         & 73.2\%                & 74.9\%                         & 74.8\%                   & 67.4\%                         & 70.0\%                  \\ \hline
		Ignite 					& 77.5\%						& 76.1\%					& 75.8\%						& 76.1\%					& 68.7\%						& 69.1\%\\ \hline
		Nutch                   & 73.1\%                         & 66.8\%                   & 76.3\%                         & 78.0\%                      & 62.2\%                         & 46.1\%                     \\ \hline
		OpenJPA                 & 78.3\%                         & 77.7\%                & 74.3\%                         & 70.0\%                    & 68.2\%                         & 62.1\%                        \\ \hline
		Pig   						& 70.1\%	 					&67.4\%	 					& 69.6\%						& 70.2\%					& 65.9\%						& 66.5\%\\ \hline
		Struts2                  & 89.3\%                         & 85.8\%                   & 87.8\%                         & 96.7\%                     & 71.2\%                         & 73.7\%                        \\ \hline
		Wicket                  & 66.6\%                         & 65.3\%                         & 72.1\%                         & 71.8\%                       & 62.7\%                         & 59.0\%                \\ \hline
		ZooKeeper 			& 69.9\%						&69.6\%						& 67.8\%						& 67.2\%					& 65.5\%						& 64.4\%\\ \hline
		Average      		  & 74.7\%						&71.6\%							&73.6\%						&73.5\%						&65.3\%						&63.5\%	 \\ \hline 
		
	\end{tabular}
\end{table}

\vspace{3mm}
\noindent\emph{\textbf{RQ2:} To what extent do unrelated software systems tend to share properties with respect to issue- and change-proneness?}
\vspace{1mm}

The results obtained in answering RQ1 showed that architectural smells  consistently impact  the issue- and change-proneness of a software system during its lifetime. In that sense, RQ2 can be considered  an extension of RQ1: we aim to understand whether architectural smells have consistent impacts across \emph{unrelated} software systems, more specifically, whether the issue- and change-proneness of a system can be accurately predicted by  models trained with data from  unrelated  systems. More deeply, this research question tries to assess whether there are  fundamentally shared traits across software systems, regardless of their developers and development processes, implementation features, application domains, underlying designs, etc.

\looseness-1
To answer this  question, instead of using 10-fold cross-validation, we selected each subject system as the test system and used its dataset as the test set; the training set was then created by combining datasets of the remaining nine systems.  
For  reference, we also built a prediction model by combining all ten systems, i.e., including the test systems.
Note that the datasets of different subject systems have different sizes; we had to resample those datasets to the same size before combining them. 

\vspace{3mm}
\noindent\emph{A. Issue-Proneness}
\vspace{1mm}

\looseness-1
Tables \ref{tab:rq2_acdc_p}, \ref{tab:rq2_arc_p}, and \ref{tab:rq2_pkg_p} summarize the precision and recall values of \emph{RQ2} experiments with regard to predicting issue-proneness under ACDC, ARC, and PKG, respectively. The left-most columns of these tables show the lists of systems. The precision and recall values are presented for three different cases: 
\begin{enumerate}
\item ``10-fold'' column -- 10-fold cross-validation on the test set. We reproduce this result from RQ1 for easy reference.
\item ``All 10'' column -- Models trained by datasets from all 10 systems, including the test set.
\item ``9 Others'' column -- Models trained by 9 other systems' datasets, not including the test set.
\end{enumerate}

\looseness-1
\noindent In total, beside the {300} issue-proneness prediction models per system that emerged from RQ1's analysis, we built and evaluated 60 additional issue-proneness  models to answer RQ2.

\def\rowswitch#1\\{} 

\begin{table}[]
	\centering
	\caption{Predicting issue-proneness -- precision~(top) and recall (bottom) under ACDC}
	\vspace{-1mm}
	\label{tab:rq2_acdc_p}
	\begin{tabular}{|l|p{1.7cm}|p{1cm}|p{1cm}|}
		\hline
		\rowcolor[HTML]{C0C0C0}
		System    & 10-fold (RQ1) & All 10 & 9 Others   \\ \hline
		Camel     & 69.9\%  & 64.8\% & 53.6\%     \\ \hline
		CXF       & 78.0\%  & 71.4\% & 66.4\%     \\ \hline
		Hadoop    & 81.2\%  & 71.1\% & 62.8\%     \\ \hline
		Ignite    & 78.9\%  & 73.9\% & 60.2\%     \\ \hline
		Nutch     & 80.8\%  & 74.9\% & 59.6\%     \\ \hline
		OpenJPA   & 71.4\%  & 68.8\% & 63.9\%     \\ \hline
		Pig       & 71.7\%  & 66.8\% & 61.4\%     \\ \hline
		Struts2   & 89.2\%  & 77.1\% & 69.1\%     \\ \hline
		Wicket    & 69.2\%  & 66.7\% & 55.0\%     \\ \hline
		ZooKeeper & 72.0\%  & 65.4\% & 56.0\%     \\ 
	\end{tabular}
	\begin{tabular}{|l|p{1.7cm}|p{1cm}|p{1cm}|}
		\hline
		\hline
		\hline
		Camel     & 68.4\%  & 57.5\% & 46.7\%     \\ \hline
		CXF       & 76.7\%  & 71.3\% & 65.7\%     \\ \hline
		Hadoop    & 80.1\%  & 69.2\% & 62.9\%     \\ \hline
		Ignite    & 78.1\%  & 73.5\% & 59.3\%     \\ \hline
		Nutch     & 71.6\%  & 68.8\% & 54.4\%     \\ \hline
		OpenJPA   & 68.3\%  & 63.0\% & 57.3\%     \\ \hline
		Pig       & 69.1\%  & 64.1\% & 58.8\%     \\ \hline
		Struts2   & 89.0\%  & 76.4\% & 68.8\%     \\ \hline
		Wicket    & 70.1\%  & 66.0\% & 54.9\%     \\ \hline
		ZooKeeper & 72.6\%  & 60.3\% & 56.9\%     \\ \hline
	\end{tabular}
\end{table}

\begin{table}[]
	\centering
	\caption{Predicting issue-proneness -- precision~(top) and recall (bottom) under ARC}
	\vspace{-1mm}
	\label{tab:rq2_arc_p}
	\begin{tabular}{|l|p{1.7cm}|p{1cm}|p{1cm}|}
		\hline
		\rowcolor[HTML]{C0C0C0}
		System    & 10-fold (RQ1) & All 10 & 9 Others   \\ \hline
		Camel     & 70.8\%  & 64.9\% & 59.7\%     \\ \hline
		CXF       & 68.9\%  & 55.2\% & 49.0\%     \\ \hline
		Hadoop    & 76.6\%  & 67.6\% & 59.6\%     \\ \hline
		Ignite    & 78.9\%  & 66.9\% & 62.3\%     \\ \hline
		Nutch     & 82.5\%  & 64.6\% & 62.3\%     \\ \hline
		OpenJPA   & 74.5\%  & 66.9\% & 63.9\%     \\ \hline
		Pig       & 71.3\%  & 62.1\% & 61.7\%     \\ \hline
		Struts2   & 95.0\%  & 76.1\% & 63.8\%     \\ \hline
		Wicket    & 76.7\%  & 63.3\% & 62.0\%     \\ \hline
		ZooKeeper & 70.8\%  & 66.3\% & 50.4\%     \\ 
	\end{tabular}
	\begin{tabular}{|l|p{1.7cm}|p{1cm}|p{1cm}|}
		\hline
		\hline
		\hline
		Camel     & 67.0\%  & 59.4\% & 48.5\%      \\ \hline
		CXF       & 68.3\%  & 62.3\% & 54.5\%      \\ \hline
		Hadoop    & 76.6\%  & 67.4\% & 59.4\%      \\ \hline
		Ignite    & 79.1\%  & 66.5\% & 61.6\%      \\ \hline
		Nutch     & 82.7\%  & 58.1\% & 53.9\%      \\ \hline
		OpenJPA   & 73.2\%  & 65.5\% & 62.0\%      \\ \hline
		Pig       & 71.1\%  & 62.5\% & 61.1\%      \\ \hline
		Struts2   & 94.8\%  & 75.7\% & 63.7\%      \\ \hline
		Wicket    & 77.1\%  & 65.3\% & 63.6\%      \\ \hline
		ZooKeeper & 69.2\%  & 67.1\% & 56.4\%      \\ \hline
	\end{tabular}
\end{table}

\begin{table}[]
	\centering
	\caption{Predicting issue-proneness -- Precision~(top) and recall (bottom) under PKG }
	\vspace{-1mm}
	\label{tab:rq2_pkg_p}
	\begin{tabular}{|l|p{1.7cm}|p{1cm}|p{1cm}|}
		\hline
		\rowcolor[HTML]{C0C0C0}
		System    & 10-fold (RQ1) & All 10 & 9 Others \\ \hline
		Camel     & 68.2\%  & 59.5\% & 46.0\%    \\ \hline
		CXF       & 64.7\%  & 62.7\% & 59.1\%    \\ \hline
		Hadoop    & 72.8\%  & 61.8\% & 50.2\%    \\ \hline
		Ignite    & 70.4\%  & 70.2\% & 62.6\%    \\ \hline
		Nutch     & 68.3\%  & 66.9\% & 51.9\%    \\ \hline
		OpenJPA   & 69.2\%  & 71.2\% & 53.1\%    \\ \hline
		Pig       & 68.6\%  & 68.0\% & 53.6\%    \\ \hline
		Struts2   & 79.1\%  & 92.4\% & 67.6\%    \\ \hline
		Wicket    & 63.7\%  & 66.1\% & 60.2\%    \\ \hline
		ZooKeeper & 68.7\%  & 66.3\% & 44.0\%    \\ 
	\end{tabular}
	\begin{tabular}{|l|p{1.7cm}|p{1cm}|p{1cm}|}
		\hline
		\hline
		\hline
		Camel     & 62.8\%  & 50.9\% & 43.5\%   \\ \hline
		CXF       & 63.8\%  & 60.0\% & 44.7\%   \\ \hline
		Hadoop    & 73.4\%  & 61.5\% & 50.3\%   \\ \hline
		Ignite    & 71.0\%  & 69.5\% & 62.3\%   \\ \hline
		Nutch     & 62.1\%  & 54.1\% & 50.9\%   \\ \hline
		OpenJPA   & 67.9\%  & 68.3\% & 39.2\%   \\ \hline
		Pig       & 69.5\%  & 68.0\% & 44.5\%   \\ \hline
		Struts2   & 78.3\%  & 92.0\% & 67.1\%   \\ \hline
		Wicket    & 65.4\%  & 66.1\% & 58.9\%   \\ \hline
		ZooKeeper & 69.4\%  & 66.8\% & 42.7\%   \\ \hline
	\end{tabular}
\end{table}

We found several consistent trends across all three architectural views. First, a prediction model built by combining data sets of multiple different software systems, even if the test system itself is included, has lower accuracy than the model built for that specific test system. This can be seen in all three Tables  \ref{tab:rq2_acdc_p}, \ref{tab:rq2_arc_p}, and \ref{tab:rq2_pkg_p}, where the ``All 10'' columns  have lower values for precision and recall than the corresponding ``10-fold'' (results from RQ1) columns. 

More interesting is the case where the test system is excluded and the model is trained on the datasets from the remaining nine systems (the ``9 others'' column). This  represents the scenario of using a  generic predictive model comprising  entirely different systems. The precision and recall values predictably decrease further across all three architectural views.
These results are reflective of the intuition that using datasets from different systems can create a more general-purpose model, but is also likely to add noise and reduce the model's ability to predict the properties of a specific system. 
Therefore, if a sufficiently large dataset for a given system is available, the system's prediction models should be trained only on that dataset. 

At the same time, it is interesting to note that the loss of accuracy between the ``10-fold'' and ``9~Others" models is relatively moderate: with few exceptions, it is on the order of 10-20\%. On the lower end, one example exception is PKG's precision for Wicket's issue-proneness (Table~\ref{tab:rq2_pkg_p}-top), where the discrepancy is only 3.5\%. On the higher end, an interesting exception are the precision and recall values obtained by ARC for Struts2 (Table~\ref{tab:rq2_arc_p}), which are both more than 30\% lower for the ``9~Others'' models. This ties to the above discussion of the limited types of smells that exist in Struts2: its uniqueness decreased the ability of other systems to predict its issue-proneness, just like it helped ensure highly accurate models when using only its own historical data.

\begin{figure}[t]
	\vspace{-4mm}
	\centering
	\subfloat[Precision]{{\includegraphics[width=4.3cm]{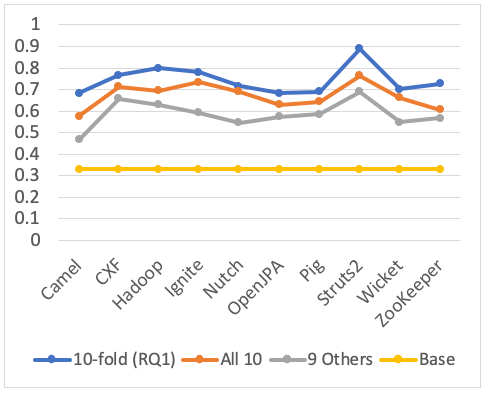} }}%
	\qquad
	\hspace{-.9cm}
	\subfloat[Recall]{{\includegraphics[width=4.3cm]{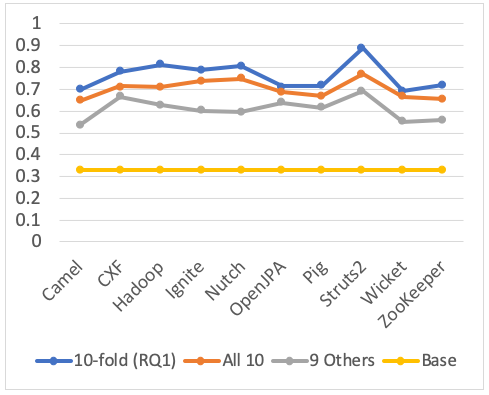} }}%
	\vspace{-1mm}
	\caption{Predicting  issue-proneness under ACDC.}
	\label{fig:compare_rq2}
\end{figure}

Figure \ref{fig:compare_rq2} shows a comparison of precision and recall between different combinations of ACDC based models. We observe that using data from ``9~Others'' systems can yield a relatively good prediction model with at least 50\% improvement compared to the baseline (0.5 vs. 0.33). In addition, the accuracy of ``All~10'' models lends support to a hypothesis that if a system has a short history of development, then including generic data can help improve predictive performance. We are currently evaluating this hypothesis more extensively.

\vspace{3mm}
\noindent\emph{B. Change-Proneness}
\vspace{1mm}

We  observed analogous trends to those discussed above in the experiments that attempt to predict the change-proneness 
using unrelated systems' datasets. We elide this data for space. 

In summary, the results of the  experiments conducted in the  context of \emph{RQ2} confirm that software systems tend to share properties with respect to issue- and change-proneness. 
The accuracy of general-purpose models is lower than that of specific models, but the gap is not prohibitive. 
Our results suggest that developers can use general-purpose models to get an overall sense of the likely issue- and change-proneness of a new software system in the early stages of its development, before sufficiently large numbers of system versions become available. Similarly, developers can use such models to predict important properties of any existing systems for which historical data is missing, spotty, or unreliable.

An interesting question is whether restricting general-purpose models to systems that are likely to share certain key characteristics can improve the models' predictive power. This is something we have not done in our current study: while the set of test systems we used share some characteristics (e.g., Java-based enterprise systems and Apache Projects), they are also inherently different systems targeting a variety of domains.
Our ongoing work is investigating whether taking into account factors such as the role of the employed development processes, off-the-shelf frameworks, system design principles and patterns, application domains, etc. can be used to increase the accuracy of the general-purpose models.

\begin{figure}[t!]
	\centering
	\includegraphics[width=7.5cm]{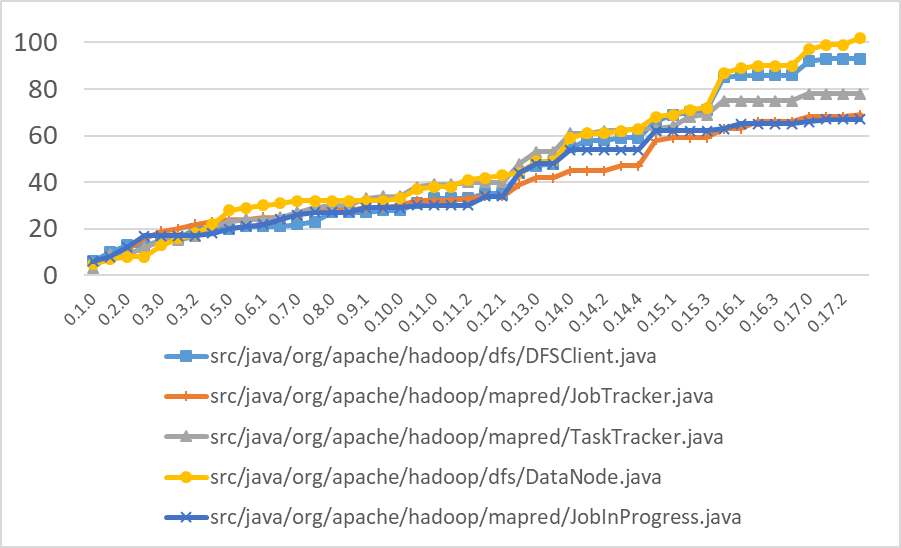}
	\includegraphics[width=7.5cm]{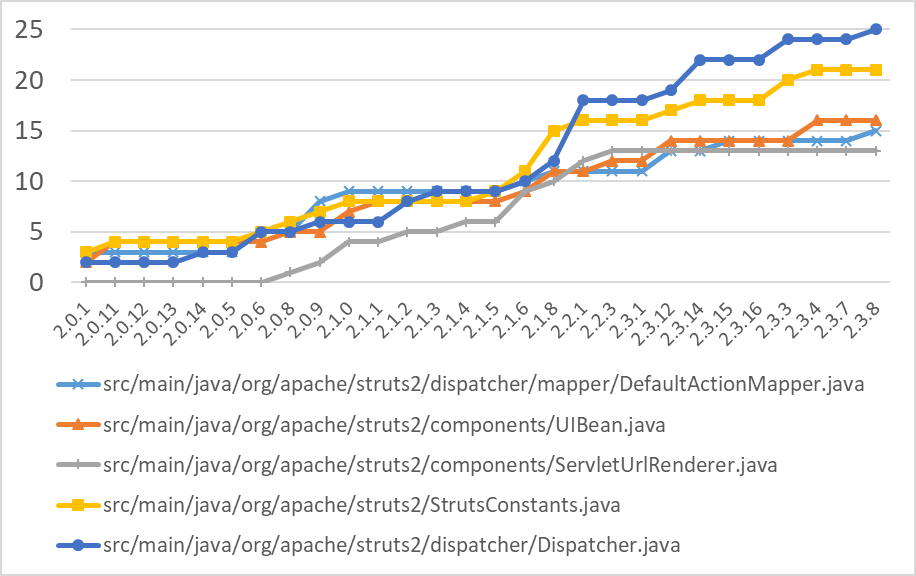}
	\vspace{-1mm}	
	\caption{Top-5 long-lived smelly files  \\ in Hadoop (top) and Struts2 (bottom).}
	\label{fig:4_struts}
\end{figure}


\looseness-1
Overall, the predictive models we developed provide developers another tool to check and maintain their software system's health and track  technical debt.
A straightforward way to identify ``unhealthy'' parts of a system is to look for \emph{long-lived smelly files}, i.e., files that have been involved in architectural smells across a large number of system versions. 
These files have a high potential to introduce  new issues.  Figure~\ref{fig:4_struts} shows examples of such files from Hadoop and Struts2. The x-axes in both plots indicate system versions, while the the y-axes indicate the numbers of smells in which each of the files is involved. 

From the collected data such as that depicted in Figure~\ref{fig:4_struts}, we have observed that long-lived smelly files are repeatedly involved in new issues during a system's lifetime. 
For example, 
\texttt{DFSClient.java} is mentioned in $\approx$2,900 Hadoop issues to date; \texttt{JobTrackers.java} is mentioned in $\approx$2,200 Hadoop issues; \texttt{Dispatcher.java} is mentioned in $\approx$670 Struts2 issues; and so on. 
We posit that stemming such trends and properly addressing the underlying problems will require considering the architectural causes of these issues.
\section{Threats to Validity}
\label{sec:thread_to_validity}

\looseness-1
The key threats to \textbf{external validity} include our subject systems. 
Most of the steps in our data gathering process are automated. However, manual intervention is required since each system has different implementation conventions. Due to the manually-intensive data gathering process, we have used data from ten subject systems in our dataset. We mitigate a possible threat stemming from the number of systems by using data from their 466 versions and evaluating 720 prediction models.

All our subject systems are Apache projects, implemented in Java, and use the Jira issue tracking system. 
The reason for this is that it helped to simplify our data gathering and analysis workflow.  In our on-going work, we are expanding our analysis beyond Apache. The diversity of the chosen systems, however, helps to reduce this threat, as does the wide adoption of Apache software, Java, and Jira. Further, all the recovery techniques and smell definitions in this paper are language-independent. %

Our study's \textbf{construct validity} is  threatened by (1)~the accuracy of the recovered architectural views, (2) the detection of architectural smells, and (3) the relevance of implementation issues. 
To mitigate the first threat, we applied three architecture recovery techniques (ACDC, ARC, and PKG) that had previously exhibited the greatest usefulness in an extensive comparative analysis of available techniques \cite{garcia2013comparative} and in a study of architectural change during system evolution \cite{leempirical, behnamghader2016large, msr2018}. The three techniques were developed independently and use  different strategies for recovering a system's architecture. 
To mitigate the second threat, we selected architectural smell types that were previously studied on a smaller scale \cite{macia2012automatically,mo2013mapping,le2016architectural, garcia2009identifying,garcia2009toward}, and were shown to be strong indicators of architectural problems. 
Finally, to mitigate the third threat, we only collected ``resolved'' and ``closed'' issues, i.e., those issues that have been independently verified and fixed by  developers. 

The primary threat to our study's \textbf{internal validity} and \textbf{conclusion validity} involves the predictability relationship between reported implementation issues and architectural smells. 
Our prediction models are built based on  significant correlations between architectural smells and implementation issues, which have been  confirmed in other work~\cite{icsa2018duc}. 
{Although correlation does not imply  causality, we have shown examples of the causal relationship's existence. Prior work has also confirmed the causality between implementation issues and architectural smells via  manual inspection \cite{Xiao:2014:DPA:2635868.2661679, 10.1007/978-3-030-00761-4_21}. }
In addition, our observations are consistent across the ten systems. 

\section{Related Work}
\label{sec:related_work}

\balance

Predicting implementation issues and code change have been widely studied research problems in software maintenance. 
The main type of implementation issues that researchers were interested early on were defects. Li et al. \cite{LI1993111} used OO metrics as predictors of software maintenance effort. Subramanyam et al.~\cite{Subramanyam1191795} also demonstrated that a set of metrics~\cite{Chidamber:1994:MSO:630808.631131} has significant implications on software defects. 
Nagappan et al.~\cite{nagappan2006mining} found a representative set of code complexity measures to determine failure-prone software entities. However,  
the metrics considered in prior work cannot prevent defects at higher abstraction levels, such as architectural problems. 

Issue prediction based on bug-fixing history is also an established area. Rahman et al. \cite{Rahman:2011:BIH:2025113.2025157} developed an algorithm that ranks files by their numbers of past changes. The algorithm helps developers find hot spots in the system that need developers' attention. There are more sophisticated methods that combine historical information and software change impact analysis to increase the efficiency and accuracy of the prediction \cite{wang2014version, hata2012bug, rahman2013and}. However, as before, these approaches do not explain higher-level defects caused by architectural decay.  

Code changes have a close connection with defects in software. Nagappan et al. \cite{nagappan2005use} used code churn to predict the defect density of software systems. Hassan et al. \cite{hassan2009predicting}  used complexity metrics based on code changes to predict  faults. Code change has been used in a number of other research efforts \cite{Xiao:2016:IQA:2884781.2884822,d2008analysing,le2016relating,leempirical} to evaluate system maintainability.

To predict code changes, Romano et al. proposed two approaches, relying on code metrics~\cite{romano2011using}  and anti-patterns~\cite{romano2012analyzing}. Xia et al.'s approach~\cite{xia2015cross} predicts a system's change-proneness using co-change information of unrelated systems. While their approach is similar to the one we employed in the context of RQ2, 
it yields relatively low accuracy. Malhotra et al. \cite{malhotra2017exploratory} used hybridized techniques to identify change-prone classes. However, their empirical study is relatively small.
Kouroshfar et al. \cite{kouroshfar2015study} do use architectural information to study the correlation between co-changes across architectural models and defects.  However, they restrict their study to cross-module changes.
\section{Conclusion}
\label{sec:conclusion}

This paper's contributions are twofold. First, we have developed an approach that can identify parts of a software system that are likely targets of future maintenance activities based on architectural characteristics as well as the change- and issue-proneness of different architectural elements. Second, we have conducted an empirical study that highlights the impact of architectural decay on ten well known open-source systems.  

We leverage  the identified correlations between symptoms of architectural decay and reported implementation issues to develop an architecture-based approach that accurately predicts a system's issue- and change-proneness. Our approach has been validated on ten existing  systems, considering 11 different types of smells under three different architectural views. This is the first study of its kind and, as such, its results can be treated as a foundation on which subsequent work should build. At the same time, the study has resulted in several important findings regarding the predictive power of architecture-based models. 

Our study confirmed that architectural smells  consistently impact a system's implementation during the system's lifecycle. In other words, the impact does not change significantly with other factors such as system size. This means that the detected architectural smells can help to accurately predict the issue-proneness and change-proneness of a system at any relevant point in time. 
In turn, such architecture-based prediction can serve as a useful tool for maintainers to recognize future problems associated with newly smell-impacted parts of the system and to plan their activities. 

As a perhaps more unexpected result, we have shown that unrelated software systems tend to share properties with respect to issue- and change-proneness. This allows developers to use general-purpose models created  with the available data from a set of existing  systems to predict the properties of  systems for which such information is missing.  
Unsurprisingly, the accuracy of such general-purpose models is lower than that of system-specific models, but not prohibitively so. Our results suggest that it is possible to develop such models sufficiently  accurately to use them as a basis of actionable advice.

\looseness-1
It is important to keep in mind that this was an initial attempt at constructing general-purpose prediction models. Our models were trained using all architectural smells and software systems  without particular prior planning. Our future work will  investigate  how to select an appropriate  set of systems to improve the accuracy of these  models. We will also explore whether further accuracy improvements can be achieved by restricting the types of architectural smells on which the models are trained.

\section{Acknowledgments}
This work is supported by the U.S. National Science
Foundation under grants 1717963, 1823354, and 1823262 and
U.S. Office of Naval Research under grant N00014-17-1-2896.

%
\bibliographystyle{abbrv}
\clearpage
\bibliography{references}  

%
%

\end{document}